 \documentclass[final,5p,times,twocolumn,square]{elsarticle}

\usepackage{amssymb}
\usepackage{lipsum}
\usepackage{lineno,hyperref}
\usepackage{lineno}
\usepackage{graphicx}
\usepackage{subcaption}
\usepackage{amsmath}
\usepackage{verbatim}
\usepackage{color}

\usepackage{mathrsfs}
\modulolinenumbers[5]

\journal{Physics Letters B}

\biboptions{numbers,sort&compress}

\begin{document}

\begin{frontmatter}



\title{Rosen-Morse potential and gravitating kinks }

\author[]{Hui Wang}
\author[]{Yuan Zhong\corref{mycorrespondingauthor}}
\cortext[mycorrespondingauthor]{Corresponding author}
\ead{zhongy@mail.xjtu.edu.cn}
\author[]{Ziqi Wang}
\address[address]{MOE Key Laboratory for Nonequilibrium Synthesis and Modulation of Condensed Matter, \\ School of Physics, Xi’an Jiaotong University, Xi’an 710049, China}

\begin{abstract}
We show that in a special type of two-dimensional dilaton-gravity-scalar model, where both the dilaton and the scalar matter fields have noncanonical kinetic terms, it is possible to construct kink solutions whose linear perturbation equation is a Schr\"odinger-like equation with Rosen-Morse potential. For this potential, eigenvalues and wave functions of the bound states, if had any, can be derived by using the standard shape invariance procedure. Depending on the values of the parameters, the stability potential can be reflective or reflectionless. There can be an arbitrary number of shape modes, but the zero mode is always absent.
\end{abstract}




\end{frontmatter}




\section{Introduction}
\label{sec:intro}

Supersymmetric quantum mechanics (SUSY QM) plays an important role in the study of exactly solvable one-dimensional quantum problems~\cite{CooperKhareSukhatme1995}. A key ingredient of SUSY QM is the so-called superpotential $\mathcal{W}(x)$, with which we can define a differential operator $\mathcal{A}=\frac{d}{dx}+\mathcal{W}$. Using $\mathcal{A}$ and its Hermitian conjugate $\mathcal{A}^{\dagger}=-\frac{d}{dx}+\mathcal{W}$, one can define two partner Hamiltonians
\begin{eqnarray}
\label{defHplus}
H_+&=&\mathcal{A}\mathcal{A}^\dagger=-\frac{d^2}{dx^2}+V_{+}(x),\\
H_-&=&\mathcal{A}^\dagger\mathcal{A}=-\frac{d^2}{dx^2}+V_{-}(x),
\end{eqnarray}
where ${V}_{\pm}=\mathcal W^2\pm \frac{d \mathcal W}{dx}$
are the partner potentials. 

It was known that for a restricted class of  $\mathcal W(x)$, the partner potentials $V_{\pm}(x)$ are related by an equation of the following form
\begin{equation}
\label{shapeinvariance}
   V_{+}(x,a_0)=V_{-}(x,a_1)+\mathcal R(a_1), 
\end{equation}
where $a_0$ is a set of parameters, $a_1$ is a function of $a_0$, and the remainder $\mathcal R(a_1)$ is independent of $x$. The partner potentials $V_{\pm}(x)$ related by Eq.~\eqref{shapeinvariance} are called \emph{shape invariant potentials} (SIP), and the corresponding Schr\"odinger-like equations are exactly solvable~\cite{Gendenshtein1983,CooperGinocchioKhare1987,DuttKhareSukhatme1988}. 

The P\"oschl-Teller II (PT II) and the Rosen-Morse (RM) potentials are two typical shape invariant potentials, and their corresponding superpotentials are
 \begin{equation}
      \label{PT2}
\mathcal{W}_{
\text{PT II}}(x,l,m)=l \tanh{x}-m\coth{x},\quad m<l,
 \end{equation}
 and
 \begin{equation}
 \label{RM}
     \mathcal{W}_{\text{RM}}(x,l,m)=l \tanh{x}+\frac{m}{l},\quad m<l^2,
 \end{equation}
where $l$ and $m$ are two real parameters. Obviously, the PT II and the RM potentials are different only when $m\neq 0$. 
The solvability of PT II and RM potentials provides profound and invaluable insights to many branches of physics, especially, the study of kink.

Kink is a type of  topological soliton solutions, which exist in many 1+1-dimensional scalar field models with degenerated vacua, and has been extensively studied in both condensed matter \cite{BishopSchneider2012} and high energy physics \cite{Vachaspati2007}. An interesting thing is that the linear perturbation equation (or stability equation) of many kink solutions, both in models with flat \cite{Kumar1987,BoyaCasahorran1989,JunkerRoy1997,AndradeMarquesMenezes2020,ZhongLiu2014,ZhongGuoFuLiu2018} and curved spacetime \cite{DeWolfeFreedmanGubserKarch2000,Giovannini2001,ZhongLiu2013,Zhong2022}, can be recast into Schr\"odinger-like equations with factorizable Hamiltonians, and therefore, can be analyzed by using the methods of SUSY QM. Some kink solutions are even related to exactly solvable shape invariant potentials.

For instance, in flat spacetime, the sine-Gordon kink and the $\phi^4$ kink are related to the regular ($m=0$) P\"oschl-Teller II potentials with parameter  $l=1$ and $2$, respectively~\cite{DashenHasslacherNeveu1974,GoldstoneJackiw1975,Rajaraman1975}, while the $\phi^6$ kink to the Rosen-Morse potential with parameter $l=3/2$ and $m=3/4$~\cite{Lohe1979}. For all these potentials, there is  a bound state with zero eigenvalue (the zero mode), but the one for the $\phi^4$ kink has one more bound state with positive eigenvalue (a shape mode). The existence of shape modes causes rich dynamical and quantum phenomena of kinks, such as the fractal structures of bounce windows~\cite{AnninosOliveiraMatzner1991,BelovaKudryavtsev1997,CampbellSchonfeldWingate1983,GoodmanHaberman2005,TakyiWeigel2016}, wobbling kink multiple scattering~\cite{CamposMohammadi2021,AlonsoIzquierdoQueirogaNunesNieto2021}, and spectral walls~\cite{AdamOlesRomanczukiewiczWereszczynski2019,EvslinHalcrowRomanczukiewiczWereszczynski2022}.

Thus, it would be interesting to ask what parent field models would generate static kink solutions with more general shape invariant stability potentials, which support more than one shape modes. In flat spacetime, this question was first addressed by Christ and Lee in 1975~\cite{ChristLee1975}, where they constructed kink models related to the regular PT II potential with arbitrary positive integer $l>0$. Relations between kink and other shape invariant potentials, such as the Rosen-Morse potential are discussed in Refs.~\cite{TrullingerFlesch1987,CasahorranNam1991a,FloresHidalgoSvaiter2002}. The one-loop quantum mass corrections for these kink solutions was studied in Refs.~\cite{BoyaCasahorran1990,AlonsoIzquierdoMateosGuilarte2012}. 

In 2022~\cite{Zhong2022a}, one of the present authors found that in a two-dimensional gravity model, it is possible to construct static kink solutions related with singular PT II potentials with parameters $m=1$ and $l>0$. Such self-gravitating kink solution can be regarded as the 2D counterpart of some 5D thick brane solutions~\cite{Stoetzel1995,Zhong2021,ZhongLiLiu2021,FengZhong2022}.
 
In the present work, we report another 2D gravitating kink model where the potential of the linear perturbation equation (or stability equation) is the Rosen-Morse potential, which reduces to the regular PT II potential as $m\to 0$. In our case with gravity, it is not easy to reconstruct the parent theory from the linear spectrum, so we will start from a general 2D dilaton-scalar-gravity model, where both the dilaton and the scalar fields can have noncanonical dynamics.

In the next section, we first derive the linear stability equation of our model, and recast it into a Schr\"odinger-like equation with factorizable Hamiltonian. We will give a condition for which the stability potential can be given analytically, once the background solution is obtained.

In Sec. \ref{sec:twinlike}, we construct a sine-Gordon type kink {solution}, and show that under some circumstances the stability potential reduces to the RM potential, or the regular PT II potential. Both potentials may have a finite number of shape modes, and the zero mode are always absent.We summary the main results in Sec.~\ref{sec:Conclusion}. 

\section{The model and linear perturbation analysis}
\label{sec:Model}
In present work, we extend the model of Ref.~\cite{ZhongGuoLiu2023,WangZhongWang2024} to the case with {noncanonical} scalar matter field, for which the action reads
\begin{equation}
 \mathcal{S}=\frac{1}{\kappa}\int d^{2}x\sqrt{-g}\left [ \varphi R+\mathcal{F}(\mathcal{X})+\kappa\mathcal{L}(X,\phi)\right ], 
 \end{equation}
where $\kappa>0$ is the gravitational coupling constant, $\mathcal{X}=-\frac{1}{2}g^{\mu\nu}\partial_{\mu}\varphi\partial_{\nu}\varphi$ and $X=-\frac{1}{2}g^{\mu\nu}\partial_{\mu}\phi\partial_{\nu}\phi$ are the kinetic terms of the dialton field and scalar field, respectively. $\mathcal{F}$ and $\mathcal{L}$ are some functions of the corresponding arguments. 

With this action, we have three dynamical equations, namely, the dilaton equation
 \begin{equation}
      \nabla^{\lambda}\left (\mathcal{F}_{\mathcal{X}}\nabla_{\lambda}\varphi\right)+R=0,
       \label{dilaton eq}
  \end{equation}
the scalar equation
\begin{equation}
       \mathcal{L}_X \nabla_\lambda \nabla^\lambda \phi+\nabla_\lambda \mathcal{L}_X \nabla^\lambda \phi+\mathcal{L}_\phi=0,
	\label{scalar eq}
 \end{equation}
and the Einstein equation
\begin{equation}
\mathcal{F}_{\mathcal{X}} \nabla_{\mu}\varphi \nabla_{\nu}\varphi-\frac{1}{2}g_{\mu\nu}
\left(-2\mathcal{F}+4\nabla_{\lambda}\nabla^{\lambda}\varphi\right)+2\nabla_{\mu}\nabla_{\nu}\varphi+\kappa T_{\mu\nu}=0,
	\label{Ein eq}
      \end{equation}
where $T_{\mu\nu}=g_{\mu \nu} \mathcal{L}+\mathcal{L}_X \nabla_\mu \phi \nabla_\nu \phi$ is the energy-momentum tensor. For simplicity, we have used subscriptions to represent derivatives of functions, for example, $\mathcal{F}_{\mathcal{X}}\equiv\frac{\partial \mathcal{F}}{\partial \mathcal{X}}$, $\mathcal{L}_{X}\equiv\frac{\partial \mathcal{L}}{\partial X}$,  and $\mathcal{L}_{\phi}\equiv\frac{\partial \mathcal{L}}{\partial \phi}$.

We focus on static kink solutions with the following metric \cite{Stoetzel1995, Zhong2021}:
 \begin{equation}
      ds^2=-e^{2A(x)}dt^2+dx^2,
      \label{metric}
  \end{equation}
 where $A(x)$ is the wrap factor. This metric can be regarded as a 2D version of the Randall-Sundrum braneworld metric \cite{RandallSundrum1999,RandallSundrum1999a}, for which the dilaton equation $\eqref{dilaton eq}$ becomes~\cite{ZhongGuoLiu2023,WangZhongWang2024}
  \begin{equation}
 \label{A1varphi}
\partial_x A=\frac{1}{2} \mathcal{F}_{\mathcal{X}} \partial_x \varphi.
 \end{equation}
After substituting Eqs.~\eqref{metric} and \eqref{A1varphi} into the Einstein equation ($\ref{Ein eq}$) we obtain
\begin{eqnarray}
 \label{SMEine eq22}
-2\partial_x^2\varphi&=&\kappa\mathcal{L}_{X}(\partial_x\phi)^2,\\ 
 \label{SMEine eq12}
-\mathcal{F}&=&\kappa(\mathcal{L}+\mathcal{L}_{X}(\partial_x\phi)^2).
 \end{eqnarray}
Finally, we have the scalar field equation:
\begin{equation}
\label{Mscalar}
\mathcal{L}_{X} \left(\partial_x A \partial_x\phi +\partial_x^2\phi\right)+\partial_x \mathcal{L}_{X} \partial_x\phi+\mathcal{L}_{\phi}=0.
 \end{equation}
Note that only three of the four equations \eqref{A1varphi}-\eqref{Mscalar} are independent.

Since our goal is to find kink solutions  with exactly solvable linear perturbation equation, it would be useful to conduct a general analysis on the linear stability for an arbitrary static solution.

To start with, we first introduce a coordinate transformation~\cite{Zhong2021,ZhongLiLiu2021}
\begin{equation}
\label{coortrans1}
r(x) \equiv \int_0^x e^{-A(\tilde{x})} d \tilde{x}.
\end{equation}
In this coordinates, the metric becomes conformally flat:
\begin{equation}
\label{flatmetric}
ds^2=e^{2A(r)}\left(-dt^2+dr^2\right),
\end{equation}
and the field equations~\eqref{A1varphi}-\eqref{Mscalar} become
 \begin{eqnarray}
 \label{EqDilaton}
A'&=&\frac{1}{2} \mathcal{F}_{\mathcal{X}}  \varphi',\\
   \label{r-E12}
-2 \varphi ''&=& \kappa  \mathcal{L}_{X} \phi '^2 -\mathcal{F}_\mathcal{X} \varphi '^2,\\
       \label{r-E2 Sim}
- \mathcal{F}&=& \kappa \left( \mathcal{L}+  e^{-2 A}\mathcal{L}_{X}\phi '^2\right),
\end{eqnarray}
and
  \begin{equation}
    \label{r-scalar}
 e^{-2 A}\left(\mathcal{L}_{X}' \phi '+ \mathcal{L}_{X} \phi ''\right)+\mathcal{L}_{\phi}=0.
\end{equation}
respectively. Here, we have used primes to denote the derivatives with respect to $r$.

Now assuming that $\left \{\varphi(r),\phi(r),g_{\mu\nu}(r)\right \}$ is an arbitrary solution of Eqs.~\eqref{EqDilaton}-\eqref{r-scalar}, and $\left \{ \delta\varphi(r,t),\delta\phi(r,t),\delta g_{\mu\nu}(r,t) \right \} $ is a small perturbation around this solution. Following Refs.~\cite{Zhong2021,ZhongLiLiu2021}, we define the metric perturbation as 
\begin{equation}
\begin{aligned}
\delta g_{\mu \nu}(r, t) & \equiv e^{2 A(r)} h_{\mu \nu}(r, t) \\
& =e^{2 A(r)}\left(\begin{array}{cc}
h_{00}(r, t) & \Phi(r, t) \\
\Phi(r, t) & h_{r r}(r, t)
\end{array}\right),
\end{aligned}
\end{equation}
and working in the dilaton gauge $ \delta \varphi=0$.

By expanding the Einstein equation \eqref{Ein eq} around the background solution and keep only the linear terms of the perturbations, we obtain two important relations:
      \begin{equation}
      \label{hrr}
 \varphi' h_{rr}=\kappa \mathcal{L}_{X}\phi'\delta \phi,
      \end{equation}
 and
  \begin{equation}
    \label{Xi} 
      \Xi= \kappa\mathcal{L}_{X}\frac{\phi'}{\varphi'} 
      \left[\gamma\delta\phi'+\delta\phi\left(\frac{\gamma\varphi''}{\varphi'}-\frac{\gamma\phi''}{\phi'}-\mathcal{F}_{\mathcal{XX}}\mathcal{X}\varphi'\right)
    \right],          
      \end{equation}
where we have defined
       \begin{equation}
    \label{gammadef}
   \Xi\equiv 2 \dot{\Phi} -h^{\prime}_{00},\quad
      \gamma\equiv 1 +\frac{2 \mathcal{L}_{XX} X}{\mathcal{L}_{X}}.
      \end{equation}

Another independent perturbation equation can be obtained by linearizing Eq.~\eqref{scalar eq}. After eliminate $h_{rr}$ and $\Xi$ by using Eqs.~\eqref{hrr} and \eqref{Xi}, we finally obtain an equation of $\delta\phi$\footnote{The derivation of Eq.~\eqref{pertur} is a bit complicated, where a series of identities have been used. More details can be found in the Mathematica notebook ``RM potential.nb", which can be downloaded via https://github.com/zhongy2009/RM-potential.git.  }:
\begin{eqnarray}
\label{pertur}
&&-\ddot{\delta \phi }+\gamma  \delta \phi ''+\gamma  \left(\frac{\gamma '}{\gamma}+\frac{  \mathcal{L}_{X}'}{{\mathcal{L}_{X}}}\right)\delta \phi ' \nonumber\\
		&& +\gamma
	 \left[
	 	-\frac{\mathcal{L}_{X}'X'}{2\mathcal{L}_{X}X}+\frac{X'^2}{4X^2}-\frac{3\kappa\mathcal{L}_{X}X'\phi'^2}{4X\varphi'}-\frac{X''}{2X}\right.\nonumber\\
		&&\left.
		-\frac{\kappa^2\mathcal{L}_{X}^2\phi'^4}{2\varphi'^2}-\frac{\kappa\mathcal{L}_{X}'\phi'^2}{\varphi'}-\frac{\kappa\mathcal{L}_{X}\phi'\phi''}{2\varphi'}
	\right]\delta \phi \nonumber\\
		&& -\gamma '
  \left(\frac{X'}{2X}+\frac{\kappa\mathcal{L}_{X}\phi'^2}{2\varphi'}\right)\delta\phi=0,
\end{eqnarray}
where the dots over $\delta\phi$ represent derivatives with respect to $t$.
We see that the above equation is independent from the form of $\mathcal {F(X)}$. In fact, all the $\mathcal {F}$ terms are eliminated by using the following identity:
 \begin{eqnarray}
      &&\mathcal{F}_{\mathcal{X}}\left (\varphi''-\frac{\varphi'\phi''}{\phi'}-\frac{\mathcal{L}_{X}'\varphi'}{2\mathcal{L}_{X}}\right)+ \mathcal{F}_{\mathcal{XX}}\mathcal{X}\left(\varphi''-\frac{1}{2} \mathcal{F}_{\mathcal{X}}\varphi'^2\right)\nonumber\\
      &=&\frac{\varphi'''}{\varphi'}-\frac{\mathcal{L}_{X}'\varphi''}{\mathcal{L}_{X}\varphi'}-\frac{2\phi''\varphi''}{\phi'\varphi'},
     \label{Iden2}
 \end{eqnarray}
which is a consequence of the background equations \eqref{EqDilaton}-\eqref{r-E12}.

A further simplification of Eq.~\eqref{pertur} is possible if we define a new variable
\begin{equation}
G(t,r)\equiv\mathcal{L}_{X}^{1/2} \gamma^{1/4}\delta\phi(t,r),
\end{equation}
along with another coordinate transformation
 \begin{equation}
 r\to y(r)\equiv \int_0^r \gamma^{-1/2}(\tilde r)d\tilde r.
  \label{dydr}
  \end{equation}
Both definitions are valid if
\begin{equation}
\mathcal{L}_{X}>0,\quad\gamma>0.
\label{condition}
\end{equation}

In $(t, y)$-coordinates, Eq.~\eqref{pertur}  becomes a compact wave equation for $G(t,y)$:
 \begin{equation}
-\partial_t^2 G+\partial_y^2 G-V_{\text{eff}}(y)G=0,
\label{wave fun}
\end{equation}
where
\begin{equation}
V_{\text{eff}}(y)\equiv\frac{\partial_y^2 f}{f},\quad f(y)\equiv \mathcal{L}_{X}^{1/2}\gamma^{1/4}\frac{\partial_{y}\phi}{\partial_{y}\varphi}.
\end{equation}
The form of $V_{\text{eff}}(y)$ is so special that if we conduct a mode expansion
\begin{equation}
\label{modeex}
G(t,y)=\sum_{n}e^{i\omega_{n}t}\psi_{n}(y),
\end{equation}
then Eq.~\eqref{wave fun} becomes a Schr\"odinger-like equation with factorizable Hamiltonian:
     \begin{equation}
     H\psi_n\equiv\mathcal{A}\mathcal{A}^\dagger\psi_n= -\frac{d^2 \psi_n}{dy^2}+V_{\text{eff}}(y)\psi_n=\omega_n^2 \psi_n,
     \label{S-like}
      \end{equation}
where
\begin{equation}
\mathcal{A}=\frac{d}{dy}+\mathcal{W},\quad\mathcal{A}^{\dagger}=-\frac{d}{dy}+\mathcal{W},
\end{equation}
and
\begin{equation}
 \mathcal{W}\equiv \frac{\partial_{y}f}{f}.
 \label{superW}
 \end{equation}
 
According to  SUSY QM theory~\cite{CooperKhareSukhatme1995}, such a factorized Hamiltonian usually has a semi-positive definite spectrum, in other words, $\omega_{n}^{2}\ge 0$. In this sense, one may say that an arbitrary static kink solution for models which satisfy the conditions in Eq.~\eqref{condition} is stable against linear perturbation\footnote{On the contrary, if some of the eigenvalues are  negative, the corresponding perturbation modes will evolve exponentially. In this case, we say that the background solution is unstable.}. The zero mode, if {existed}, has a wave function $\psi_0\propto f$. 

However, it is not easy to find a kink solution under the metric \eqref{metric}, such that the linear perturbation equation~\eqref{S-like} is exactly solvable. Because, in general, the coordinate transformations \eqref{coortrans1} and \eqref{dydr} are so complicate that $V_{\text{eff}}(y)$  does not even have an analytical expression, needless to say to be solved exactly.

Nevertheless, as shown in Ref.~\cite{Zhong2022a} in a simpler model (where $\mathcal{F(X)=X}$), there is a special situation when the second coordinate transformation inverses the first one. In this case
\begin{equation}
\frac{dy}{dx}=\frac{dy}{dr}\frac{dr}{dx}=\gamma^{-1/2} e^{-A}=1,
\label{dyda1}
\end{equation}
which means
\begin{equation}
\gamma=e^{-2A}.
\label{gammaA}
\end{equation}
Therefore, by properly choosing an $\mathcal L(X,\phi)$ such that the condition \eqref{gammaA} is satisfied, one would have $y=x$, and the equation \eqref{S-like} becomes
\begin{equation}
H\psi_n\equiv\mathcal{A}\mathcal{A}^\dagger\psi_n= -\frac{d^2 \psi_n}{dx^2}+V_{\text{eff}}(x)\psi_n=E_n\psi_n,
\label{S-like x} 
\end{equation}
where
\begin{equation}
    \label{VW}
E_n\equiv\omega_n^2 ,\quad V_{\text{eff}}(x)=\frac{\partial_{x}^2f}{f},\quad  f(x)\equiv\mathcal{L}_{X}^{1/2}\gamma^{1/4}\frac{\partial_{x}\phi}{\partial_{x}\varphi}.  
\end{equation}
Once we obtain a kink solution in the $x$-coordinates, we immediately obtain the expression of the stability potential $V_{\text{eff}}(x)$. 

Note that by definition, the Hamiltonian $H$ in Eq.~\eqref{S-like x} corresponds to $H_+$ of the partner Hamiltonians, and $V_{\text{eff}}(x)$ corresponds to $V_{+}(x)$ of the partner potentials. Thus, in order all the eigenvalues $E_n=E_n^{(+)}$ and the corresponding wave functions $\psi_n=\psi_n^{+}$ can be solved out exactly, the partner potentials generated by the superpotential $ \mathcal{W}= \frac{\partial_{x}f}{f}$ must be shape invariant. 

Now, let us see how to realize the this idea.

\section{A solution}
\label{sec:twinlike}

Since the condition \eqref{gammaA} is independent of function $\mathcal {F(X)}$, we can try the same K-field model used in Ref.~\cite{Zhong2022a}:
\begin{equation}
\mathcal{L}=U(\phi)(X+\frac{1}{2}W_{\phi}^2)^2+X-V(\phi),
\label{twinmodel}
\end{equation}
where $U(\phi)$, $W(\phi)$ and $V(\phi)$ are all functions of $\phi$, whose forms are to be determined. 
This Lagrangian has a special property, namely, if one assumes $W(\phi)$ to be some functions such that 
\begin{equation}
\partial_{x}\phi=W_{\phi}\equiv\frac{dW(\phi)}{d\phi},
\label{phixwphi}
 \end{equation}
 or equivalently,
\begin{equation}
\label{kineticW}
X=-\frac{1}{2}W_{\phi}^2,
 \end{equation}
 then
\begin{equation}
\mathcal{L}|  =X-V,\quad
\mathcal{L}_X|=1,\quad \mathcal{L}_{XX}|=2U(\phi),
\end{equation}
where the symbol $`` ~| ~"$ is short for  $``~ |_{X=-\frac{1}{2}W_{\phi}^2}~"$.
Consequently, the background field equations \eqref{A1varphi}-\eqref{SMEine eq12} become
\begin{eqnarray}
\label{AFw}
\partial _{x} A&=&-\frac{1}{4}\kappa ~\mathcal{F}_{\mathcal{X}} W,\\
\label{varphiw}
-2\partial_{x}\varphi&=&{\kappa} W,\\
\label{Vw}
V&=&\frac{\mathcal{F}}{\kappa}+\frac{1}{2}W_{\phi}^2.
\end{eqnarray}
and the condition in Eq.~\eqref{gammaA} becomes
\begin{equation}
U(\phi)=\frac{1}{2W_{\phi}^2}\left(1-e^{-2A(\phi) }\right).
\label{Uphi}
\end{equation}

Obviously, once $W(\phi)$ and $\mathcal {F(X)}$ are properly specified, all other variables can be derived. As an example, let us take
\begin{equation}
\label{Wphi}
W(\phi)=\text{sin}\phi-c,
\end{equation}
for which Eqs.~\eqref{phixwphi} and \eqref{varphiw} can be solved immediately: 
\begin{eqnarray}
    \label{sgphi}
\phi(x)&=&\text{arcsin}(\text{tanh}(x)),\\
\label{sgvarphi}
\varphi(x)&=&\frac{\kappa} {2}\left[cx-\text{ln}(\text{cosh}(x))\right].
\end{eqnarray} The scalar field has a kink configuration: $\phi(x\to\pm\infty)=\pm\frac\pi 2$. The dilaton field is asymmetric if the parameter $c\neq 0$.

To find out the solution of $A(x), V(\phi)$ and $U(\phi)$, we take
\begin{equation}
\mathcal{F}(\mathcal{X})=\mathcal{X}-2\alpha\sqrt{-2\mathcal{X}},
 \end{equation}
where $\alpha$ is a real parameter.  In order $\sqrt{-2\mathcal X}$ to be meaningful, we will always assume $c\geq 1$ for $\alpha\neq 0$, so that $\varphi(x)$ in Eq.~\eqref{sgvarphi} is a monotonically increasing function, and ${\partial_x\varphi}>0$. In this case, 
\begin{equation}
\mathcal{F}_\mathcal{X}=1+\frac{2\alpha }{\sqrt{-2\mathcal{X}}}=1+\frac{2\alpha }{{\partial_x\varphi}},
\end{equation}
and Eq. \eqref{AFw} becomes 
\begin{equation}
\label{Avarphi}
\partial _{x} A=-\frac{1}{4}\mathcal{\kappa } W+\alpha,
\end{equation}
which after an integration gives
\begin{equation}
A(\phi)=-\frac{\kappa}{4}\int \frac{W}{W_{\phi}}d\phi+\alpha\int\frac{1}{W_{\phi}}d\phi.
\label{Aphi}
\end{equation}

After inserting the $W(\phi)$ of Eq.~\eqref{Wphi} into the above two equations, one immediately obtains
\begin{equation}
 A(x)=\frac{1}{4} c \kappa x-\frac{1}{4} \kappa \ln (\cosh ( x))+\alpha  x,
\end{equation}
and 
\begin{equation}
 A(\phi)=\frac{1}{4} \left[(4 \alpha +c \kappa ) \rm{arctanh}(\sin \phi )+\kappa  \ln (\cos \phi )\right].
\end{equation}
Then from Eq.~($\ref{Uphi}$) we get
\begin{equation}
    U(\phi)=\frac{\sec ^2\phi }{2}-\frac{1}{2} \sec ^{\frac{\kappa }{2}+2}\phi \cdot e^{-\frac{1}{2} (4 \alpha +c \kappa ) \rm{arctanh}(\sin \phi )}.
\end{equation}
The expression for the scalar potential can be derived from Eq.~\eqref{Vw}:
\begin{eqnarray}
\label{sgV}
V(\phi)=-\alpha  \sqrt{(c-\sin \phi )^2}-\frac{\kappa}{8} (c-\sin \phi )^2+\frac{\cos ^2\phi }{2}.
\end{eqnarray}

With the solutions of the background field equations, we can now analyze the linear perturbation spectrum. Since we only care about shape invariant potentials, all we need is to derive the expression of the superpotential  $\mathcal{W}\equiv \frac{\partial_{x}f}{f}$, where the function $f(x)$ is defined in Eq.~\eqref{VW}. After inserting our solution, we obtain
\begin{equation}
f(x)=\frac{\text{sech}^{1-l}(x) e^{-c l x-\frac{\alpha  x}{2}}}{4 l (c-\tanh (x))}\propto \psi_0^+(x),
\end{equation}
and
\begin{equation}
\mathcal W=-\frac{\alpha }{2}-c l+\frac{\text{sech}^2(x)}{c-\tanh (x)}+(l-1) \tanh (x).
\end{equation}
Here we have defined $l\equiv \kappa/8>0$ \footnote{One may note that this a little definition is different from the one of Ref.~\cite{Zhong2022a}. In fact, there is an error in the Eq.~(4.6) of Ref.~\cite{Zhong2022a}, where the factor $\cosh ^{\frac{\kappa v^2}{2}}(k x)$ should be $\cosh ^{\frac{\kappa v^2}{8}}(k x)$. So in Ref.~\cite{Zhong2022a}, one should have defined $l\equiv\kappa v^2/8$. Fortunately, this error does not affect the main results.}.  To obtain shape invariant partner potentials we need to impose some constraints on the parameters, and there are at least two interesting cases.

\subsection*{Case 1: $\alpha=0$ and $c=0$}
This corresponds to the case with $\mathcal{F(X)=X}$, $W(\phi)=\sin\phi$ and has been studied in Ref.~\cite{Zhong2022a}. In this case, the superpotential $\mathcal W(x)$ reduces to the one for the singular P\"oschl-Teller II potentials with parameter $m=1$:
\begin{equation}
\mathcal W=l \tanh (x)-\coth (x).
\label{singularW}
\end{equation}
The spectra of the bound states for the partner Hamiltonian $H_\pm$ are~\cite{CasahorranNam1991,Zhong2022a}
\begin{equation}
E_{n}^{(+)}=(l-1)^2-(l-2 n-3)^2, \quad n=0,1,2, \cdots<\frac{l-3}{2},
\end{equation}
and
\begin{equation}
E_{n}^{(-)}=(l-1)^2-(l-n)^2, \quad n=0,1,2, \cdots<l, \quad l>1.
\end{equation}
We see that when $l>3$, $H_+$ can have some bound states with positive eigenvalues, while $H_-$ always have a bound state with negative eigenvalue $E_0^{(-)}=1-2l$, along with a zero mode $E_1^{(-)}=0$.  One can also show that the eigenvalues of the partner Hamiltonian are only partly degenerated. For example, when $l=6$, we have $E_0^{(+)}=E_3^{(-)}=16$, $E_1^{(+)}=E_5^{(-)}=24$, and all other bound states of $H_-$ have no correspondence in $H_+$. The reason behind these abnormal properties is that one of the partner potentials generated by the superpotential \eqref{singularW} is singular, and the supersymmetry of the system is broken.

In the present model, we have yet another possibility where the supersymmetry is unbroken~\cite{CasahorranNam1991}.

\begin{figure*}[!ht]
\centering
\includegraphics[width=1\textwidth ]{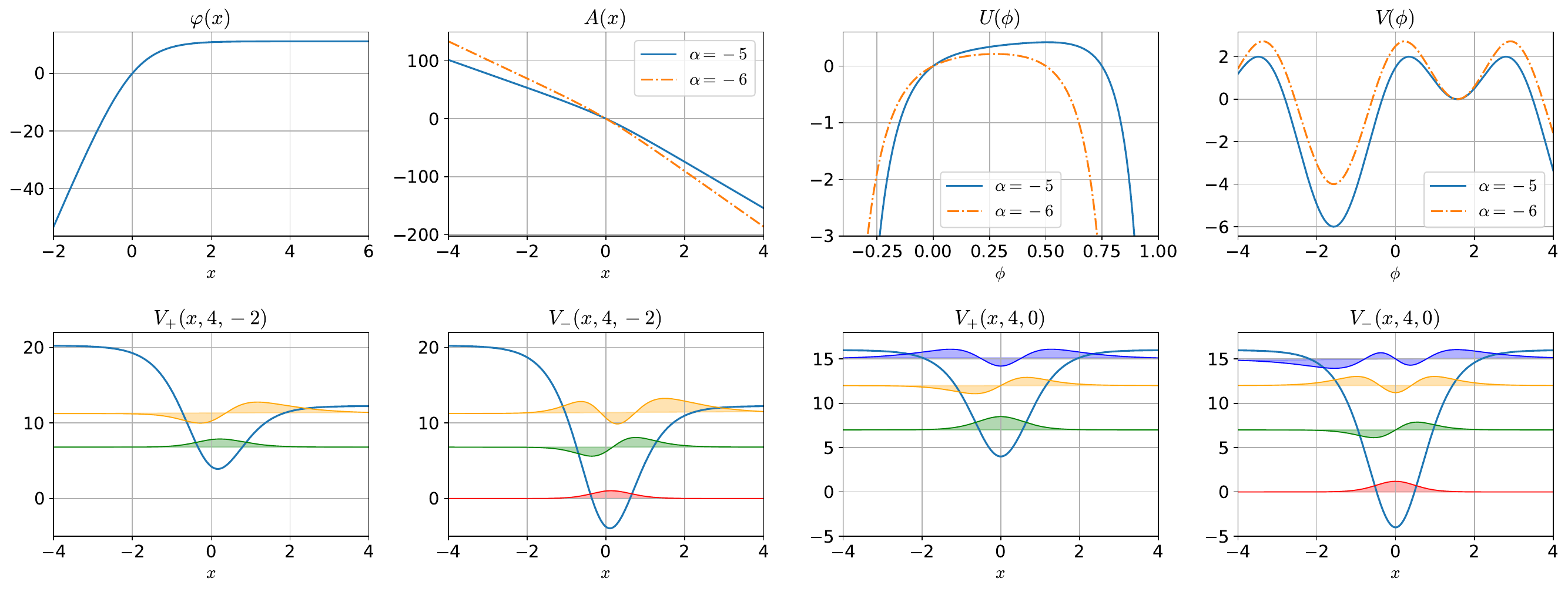}
\caption{Plots of the background solutions $\varphi(x)$, $A(x)$, $U(\phi)$, $V(\phi)$, and the partner potentials $V_\pm(x,l,m)$ which are related to the linear perturbation problem. We have set $c=1$, $l=4$, $\alpha=-5$ and $-6$.}
\label{fig1}
\end{figure*} 

\subsection*{Case 2: c=1}
In this case, the superpotential becomes the one for the Rosen-Morse potentials:
\begin{equation}
\mathcal{W}(x,l,\alpha)=l \tanh (x)+m/l,
\end{equation}
where $m\equiv l(1-l-\frac{\alpha}{2})$. The partner potentials are
\begin{equation}
V_\pm(x,l,m)=\left(\frac{m}{l}+l \tanh (x)\right)^2\pm l~ \text{sech}^2(x).
\label{V_RM}
\end{equation}
Both of the partner potentials are regular in this case.
We see that for $l>0$, $V_+(x,l,m)$ is non-negative everywhere, so there is no normalizable zero mode, and the spectrum is positive definite. Moreover, it is easy to verify that $V_\pm(x,l,m)$ are shape invariant in the following sense:
\begin{eqnarray}
\label{shape}
V_{+}(x,l,m)&=&V_{-}(x,l-1,m)\nonumber\\
&+&l^2-(l-1)^2+m^2\left[\frac{1}{ l^2}-\frac{1}{(l-1)^2}\right].
\end{eqnarray}
Note that $V_{-}(x,l-1,m)=V_{-}(x,l-1,m(l,\alpha))$ is obtained from $V_{-}(x,l,m(l,\alpha))$ by replacing $l\to l-1$ and $\alpha \to 4+{\alpha  l}({l-1)^{-1}}$. Under this replacement, $m$ does not change.

In order to use the standard shape invariant procedure to determine the spectrum of $V_+(x,l,m)$, which corresponds to the effective potential $V_{\rm{eff}}(x)$ in the linear perturbation equation \eqref{S-like x}, we need to properly confine the range of the parameters to ensure that we are in a system with unbroken SUSY.
 
The unbroken of SUSY means at least one of the zero mode $\psi_0^{(+)}\propto f$ or $\psi_0^{(-)}\propto 1/f$ is normalizable~\cite{GangopadhyayaMallowRasinariu2017}. But as we have stated above, $\psi_0^{(+)}$ is not normalizable, no matter what the parameters are. Thus, the only possibility for unbroken SUSY is the normalization of $\psi_0^{(-)}$, which requires $\mathcal{W}(x\to+\infty)>0$ and $\mathcal{W}(x\to-\infty)<0$~\cite{GangopadhyayaMallowRasinariu2017}, or equivalently,
\begin{equation}
 -l^2<m<l^2 \quad \Longleftrightarrow \quad 2-4l<\alpha<2 .
\end{equation}
With the shape invariance relation \eqref{shape} and the above condition for unbroken SUSY, the eigenvalues $E_n^{(\pm)}$ of the partner Hamiltonians $H_\pm$ can be derived (see Ref.~\cite{DuttKhareSukhatme1988} for details). For both of the partner Hamiltonians, there is a continuous spectrum which starts from $\left(l-{| m | }/ {l}\right)^2$, and a discrete spectrum~\cite{DuttKhareSukhatme1988} :
\begin{eqnarray}
\label{eigenvalue}
    E_{n}^{(-)}&=&l^2-(l-n)^2+m^2\left[\frac{1}{l^2}-\frac{1}{(l-n)^2}\right],  \\
    E_{n}^{(+)}&=&E_{n+1}^{(-)},
\end{eqnarray}
where $n=0,1,2,3,\dots <l-\sqrt{\left | m \right | }$ is a label for the bound states.  We see that the spectrum of $H_+$ is identical to the one of $H_-$, except the ground state.

Because the bound states largely impact the dynamical and quantum properties of a kink, it is interesting to see the case where $V_+(x)=V_{\rm{eff}}(x)$ has some bound states. 
In order  $V_+(x)$ has \emph{at least} $n$ bound states, we need $n <l-\sqrt{\left | m \right | }$, which means that
 \begin{equation}
l>n,\quad -4 l+2+2 n(2-\frac{ n}{l})<\alpha<2-2 n(2-\frac{ n}{l}).
\end{equation}
For example, if we chose $n=2$ and  $l=4$, then $\alpha$ must lie in the interval $-8<\alpha<-4$. 

Let us first take $\alpha=-5$, which yields $m=-2$. In this case, the continuous spectrum starts at $E=12.25$, and $V_{-}(x, 4,-2)$ has three bound states:
\begin{eqnarray}
E^{(-)}_0=0, && \psi^-_0(x)\propto e^{x/2} \text{sech}^4(x),\\
E^{(-)}_1=\frac{245}{36}, && \psi^-_1(x)\propto e^{\frac{2 x}{3}} (6 \tanh (x)-1) \text{sech}^3(x),\\
E^{(-)}_2=\frac{45}{4},&& \psi^-_2(x)\propto e^x  \text{sech}^4(x)\nonumber\\
&&\quad\quad~~\times(7 e^{-2 x}+12 \sinh ^2(x)-9),
\end{eqnarray}
while $V_{+}(x, 4,-2)$ has two:
\begin{eqnarray}
E^{(+)}_0=\frac{245}{36}, && \psi^+_0(x)\propto e^{\frac{2 x}{3}} \text{sech}^3(x),\\
E^{(+)}_1=\frac{45}{4},&& \psi^+_1(x)\propto e^x (3 \tanh (x)-1) \text{sech}^2(x).
\end{eqnarray}
Note that here and below, the subscript $`` 0 "$ in $E_0^{(+)}$ and $\psi_0^{+}$ does not represent the zero mode, but the ground state. The wave functions are derived  via the standard shape invariant procedure \cite{DuttKhareSukhatme1988}:  
\begin{align*}
\psi^+_0(x,l,m)&\propto \mathcal{A}(x,l,m)\psi^{-}_{1}(x,l,m)\\
&\propto \mathcal{A}(x,l,m)\mathcal{A}^{\dagger}(x,l,m)\psi^{-}_{0}(x,l-1,m),\\
    \psi^+_1(x,l,m)&\propto \mathcal{A}(x,l,m)\psi^{-}_{2}(x,l,m) \\
    \propto \mathcal{A}(x,l,m)&\mathcal{A}^{\dagger}(x,l,m)\mathcal{A}^{\dagger}(x,l-1,m)\psi^{-}_{0}(x,l-2,m),\\
   & \vdots
\end{align*}
where $\psi^{-}_0(x,l,m)\propto 1/f \propto e^{-\frac{m x}{l}} \text{sech}^l(x)$. More bound states can be obtained in the same manner, provided that $l$ is large enough.

Now, let us turn to a special case where $\alpha=2(1-l)$. In this case, we have $m=0$ and the Rosen-Morse potentials reduce to the reflectionless and symmetric P\"oschl-Teller II potentials: 
\begin{equation}
V_{\pm}(x,l,0)=l^2-l(l\mp1)\text{sech}^2(x).
\end{equation}
Now, the continuous spectra start from $l^2$, and the discrete spectra are $E_{n}^{(-)}=l^2-(l-n)^2$,  $E_{n}^{(+)}=E_{n+1}^{(-)}$, $n=0,1,2,\cdots <l$. Note that $V_{-}(x,2,0)$ and $V_{-}(x,1,0)$ are nothing but the potentials in the linear perturbation equation of $\phi^4$ and sine-Gordon kinks in flat space. In our case with gravity, $V_{\rm{eff}}(x)$ corresponds to $V_{+}(x,l,0)$, which has no normalizable zero mode, but can have at least $n$ bound states, if $l>n$. 

To compare with the previous example, let us take $l=4$ again, but with $\alpha=-6$ this time, so that $m=0$. The bound states of $V_{-}(x,4,0)$ and $V_{+}(x,4,0)$ are
\begin{eqnarray}
E^{(-)}_0=0,&& \psi^-_0(x)\propto \text{sech}^4(x),\\
E^{(-)}_1=7,&& \psi^-_1(x)\propto \tanh (x) \text{sech}^3(x),\\
E^{(-)}_2=12,&& \psi^-_2(x)\propto \text{sech}^2(x) (6-7 \text{sech}^2(x)),\\
E^{(-)}_3=15,&& \psi^-_3(x)\propto \tanh(x)\text{sech}(x)(4-7\text{sech}^2(x)).
\end{eqnarray}
and
\begin{eqnarray}
E^{(+)}_0=7, &&  \psi^+_0(x)\propto \text{sech}^3(x),\\
\label{Eq81}
E^{(+)}_1=12,&& \psi^+_1(x)\propto \tanh (x) \text{sech}^2(x),\\
E^{(+)}_2=15,&& \psi^+_2(x)\propto \text{sech}(x)(4-5\text{sech}^2(x)),
\end{eqnarray}
respectively. Now we have one more bound state, and the wave functions now become either odd or even functions.

 The shapes of the background solutions $\varphi(x)$, $A(x)$, $U(\phi)$, $V(\phi)$, and  the partner potentials $V_\pm(x)$ for $c=1,l=4$, $\alpha=-5$ and $-6$ can be found in Fig.~\ref{fig1}. 

\section{Conclusion}
\label{sec:Conclusion}

In this work, we generalized the 2D gravity model of Refs.~\cite{ZhongGuoLiu2023,WangZhongWang2024} to the case with noncanonical scalar matter field $\mathcal L(X,\phi)$. We first gave a general linear perturbation analysis for an arbitrary static solution of the model, and showed that the stability equation can be recasted into a Schr\"odinger-like equation with factorizable Hamiltonian. According to the knowledge of SUSY QM, a Hamiltonian of this type usually has positive semi-definite spectrum, which ensures the stability of the solution. 

Then, by using the first-order formalism of the field equations, we constructed an exact gravitating kink solution with three tunable parameters $c, l$ and $\alpha$. Here $c$ is a dimensionless model parameter, $l\equiv\kappa/8$ is a rescaling of the gravitational coupling constant, and $\alpha$ is the coefficient of the noncanonical part of  the dilaton kinetic term, which is chosen to be $\mathcal{F(X)}=\mathcal{X}-2\alpha\sqrt{-2\mathcal{X}}$ in the present model. 

When $\alpha=0=c$, our solution reduces to the one obtained in Ref.~\cite{Zhong2022a}, for which the stability potential is a subclass of the singular PT II potential (with the parameter $m=1$). In this case, there can be some shape  modes if $l>3$.

If  $\alpha\neq0$, we require $c\geq 1$ so that $\sqrt{-2\mathcal{X}}$ is well defined. An interesting observation of the present work is that when $c=1$, the partner potentials related to the linear stability problem reduce to the shape invariant Rosen-Morse potential, which is regular and exactly solvable. We showed that by tuning the values of $l$ and $\alpha$, there can be a finite number of shape modes in the linear spectrum. Especially, if $\alpha=2(1-l)$, the asymmetric Rosen-Morse potential becomes the symmetric and reflectionless PT II potential.

It would be very interesting to explore the quantum and dynamical properties of our solution, by following the methods in Refs.~\cite{OgundipeEvslinZhangGuo2024,OgundipeEvslin2024,EvslinLiu2024,EvslinLiu2024a,EvslinLiu2023,EvslinRoystonZhang2023,Evslin2022,Evslin2021,GuoEvslin2020,Evslin2019}. We would leave these issues to our future works.

\section*{Acknowledgments}
This work was supported by the National Natural Science Foundation of China (Grant number 12175169).

\section*{Declaration of competing interest}
The authors declare that they have no known competing financial interests or personal relationships that could have appeared to influence the work reported in this paper.


\begin{thebibliography}{10}
\expandafter\ifx\csname url\endcsname\relax
  \def\url#1{\texttt{#1}}\fi
\expandafter\ifx\csname urlprefix\endcsname\relax\def\urlprefix{URL }\fi
\expandafter\ifx\csname href\endcsname\relax
  \def\href#1#2{#2} \def\path#1{#1}\fi

\bibitem{CooperKhareSukhatme1995}
F.~Cooper, A.~Khare, U.~Sukhatme, {Supersymmetry and quantum mechanics}, Phys.
  Rept. 251 (1995) 267--385.
\newblock \href {http://arxiv.org/abs/hep-th/9405029}
  {\path{arXiv:hep-th/9405029}}, \href
  {https://doi.org/10.1016/0370-1573(94)00080-M}
  {\path{doi:10.1016/0370-1573(94)00080-M}}.

\bibitem{Gendenshtein1983}
L.~E. Gendenshtein, {Derivation of Exact Spectra of the Schrodinger Equation by
  Means of Supersymmetry}, JETP Lett. 38 (1983) 356--359.

\bibitem{CooperGinocchioKhare1987}
F.~Cooper, J.~N. Ginocchio, A.~Khare, {Relationship Between Supersymmetry and
  Solvable Potentials}, Phys. Rev. D 36 (1987) 2458--2473.
\newblock \href {https://doi.org/10.1103/PhysRevD.36.2458}
  {\path{doi:10.1103/PhysRevD.36.2458}}.

\bibitem{DuttKhareSukhatme1988}
R.~Dutt, A.~Khare, U.~P. Sukhatme, {Supersymmetry, Shape Invariance and Exactly
  Solvable Potentials}, Am. J. Phys. 56 (1988) 163--168.
\newblock \href {https://doi.org/10.1119/1.15697} {\path{doi:10.1119/1.15697}}.

\bibitem{BishopSchneider2012}
A.~R. Bishop, T.~Schneider, Solitons and Condensed Matter Physics: Proceedings
  of the Symposium on Nonlinear (Soliton) Structure and Dynamics in Condensed
  Matter, Oxford, England, June 27--29, 1978, Vol.~8, Springer Science \&
  Business Media, 2012.

\bibitem{Vachaspati2007}
T.~Vachaspati, {Kinks and Domain Walls : An Introduction to Classical and
  Quantum Solitons}, Oxford University Press, 2007.
\newblock \href {https://doi.org/10.1017/9781009290456}
  {\path{doi:10.1017/9781009290456}}.

\bibitem{Kumar1987}
C.~N. Kumar, \href{https://dx.doi.org/10.1088/0305-4470/20/15/051}{Isospectral
  hamiltonians: generation of the soliton profile}, Journal of Physics A:
  Mathematical and General 20~(15) (1987) 5397.
\newblock \href {https://doi.org/10.1088/0305-4470/20/15/051}
  {\path{doi:10.1088/0305-4470/20/15/051}}.
\newline\urlprefix\url{https://dx.doi.org/10.1088/0305-4470/20/15/051}

\bibitem{BoyaCasahorran1989}
L.~J. Boya, J.~Casahorran, {General scalar bidimensional models including
  kinks}, Annals Phys. 196 (1989) 361.
\newblock \href {https://doi.org/10.1016/0003-4916(89)90182-6}
  {\path{doi:10.1016/0003-4916(89)90182-6}}.

\bibitem{JunkerRoy1997}
G.~Junker, P.~Roy, Construction of (1+1)-dimensional field models with exactly
  solvable fluctuation equations about classical finite-energy configurations,
  Annals of Physics 256~(2) (1997) 302--319.
\newblock \href {https://doi.org/https://doi.org/10.1006/aphy.1997.5681}
  {\path{doi:https://doi.org/10.1006/aphy.1997.5681}}.

\bibitem{AndradeMarquesMenezes2020}
I.~Andrade, M.~A. Marques, R.~Menezes, {Stability of kinklike structures in
  generalized models}, Nucl. Phys. B 951 (2020) 114883.
\newblock \href {http://arxiv.org/abs/1906.05662} {\path{arXiv:1906.05662}},
  \href {https://doi.org/10.1016/j.nuclphysb.2019.114883}
  {\path{doi:10.1016/j.nuclphysb.2019.114883}}.

\bibitem{ZhongLiu2014}
Y.~Zhong, Y.-X. Liu, {$K$-field kinks: stability, exact solutions and new
  features}, JHEP 10 (2014) 041.
\newblock \href {http://arxiv.org/abs/1408.4511} {\path{arXiv:1408.4511}},
  \href {https://doi.org/10.1007/JHEP10(2014)041}
  {\path{doi:10.1007/JHEP10(2014)041}}.

\bibitem{ZhongGuoFuLiu2018}
Y.~Zhong, R.-Z. Guo, C.-E. Fu, Y.-X. Liu, {Kinks in higher derivative scalar
  field theory}, Phys. Lett. B 782 (2018) 346--352.
\newblock \href {http://arxiv.org/abs/1804.02611} {\path{arXiv:1804.02611}},
  \href {https://doi.org/10.1016/j.physletb.2018.05.048}
  {\path{doi:10.1016/j.physletb.2018.05.048}}.

\bibitem{DeWolfeFreedmanGubserKarch2000}
O.~DeWolfe, D.~Z. Freedman, S.~S. Gubser, A.~Karch, {Modeling the
  fifth-dimension with scalars and gravity}, Phys. Rev. D 62 (2000) 046008.
\newblock \href {http://arxiv.org/abs/hep-th/9909134}
  {\path{arXiv:hep-th/9909134}}, \href
  {https://doi.org/10.1103/PhysRevD.62.046008}
  {\path{doi:10.1103/PhysRevD.62.046008}}.

\bibitem{Giovannini2001}
M.~Giovannini, {Gauge invariant fluctuations of scalar branes}, Phys. Rev. D 64
  (2001) 064023.
\newblock \href {http://arxiv.org/abs/hep-th/0106041}
  {\path{arXiv:hep-th/0106041}}, \href
  {https://doi.org/10.1103/PhysRevD.64.064023}
  {\path{doi:10.1103/PhysRevD.64.064023}}.

\bibitem{ZhongLiu2013}
Y.~Zhong, Y.-X. Liu, {Linearization of thick K-branes}, Phys. Rev. D 88~(2)
  (2013) 024017.
\newblock \href {http://arxiv.org/abs/1212.1871} {\path{arXiv:1212.1871}},
  \href {https://doi.org/10.1103/PhysRevD.88.024017}
  {\path{doi:10.1103/PhysRevD.88.024017}}.

\bibitem{Zhong2022}
Y.~Zhong, {Normal modes for two-dimensional gravitating kinks}, Phys. Lett. B
  827 (2022) 136947.
\newblock \href {http://arxiv.org/abs/2112.08683} {\path{arXiv:2112.08683}},
  \href {https://doi.org/10.1016/j.physletb.2022.136947}
  {\path{doi:10.1016/j.physletb.2022.136947}}.

\bibitem{DashenHasslacherNeveu1974}
R.~F. Dashen, B.~Hasslacher, A.~Neveu, {Nonperturbative Methods and Extended
  Hadron Models in Field Theory 2. Two-Dimensional Models and Extended
  Hadrons}, Phys. Rev. D 10 (1974) 4130--4138.
\newblock \href {https://doi.org/10.1103/PhysRevD.10.4130}
  {\path{doi:10.1103/PhysRevD.10.4130}}.

\bibitem{GoldstoneJackiw1975}
J.~Goldstone, R.~Jackiw, {Quantization of Nonlinear Waves}, Phys. Rev. D 11
  (1975) 1486--1498.
\newblock \href {https://doi.org/10.1103/PhysRevD.11.1486}
  {\path{doi:10.1103/PhysRevD.11.1486}}.

\bibitem{Rajaraman1975}
R.~Rajaraman, {Some Nonperturbative Semiclassical Methods in Quantum Field
  Theory: A Pedagogical Review}, Phys. Rept. 21 (1975) 227--313.
\newblock \href {https://doi.org/10.1016/0370-1573(75)90016-2}
  {\path{doi:10.1016/0370-1573(75)90016-2}}.

\bibitem{Lohe1979}
M.~A. Lohe, {Soliton Structures in $P (\phi$) in Two-dimensions}, Phys. Rev. D
  20 (1979) 3120.
\newblock \href {https://doi.org/10.1103/PhysRevD.20.3120}
  {\path{doi:10.1103/PhysRevD.20.3120}}.

\bibitem{AnninosOliveiraMatzner1991}
P.~Anninos, S.~Oliveira, R.~A. Matzner, {Fractal structure in the scalar
  $\lambda (\phi^2-1)^2$ theory}, Phys. Rev. D 44 (1991) 1147--1160.
\newblock \href {https://doi.org/10.1103/PhysRevD.44.1147}
  {\path{doi:10.1103/PhysRevD.44.1147}}.

\bibitem{BelovaKudryavtsev1997}
T.~I. Belova, A.~E. Kudryavtsev, {Solitons and their interactions in classical
  field theory}, Phys. Usp. 40 (1997) 359--386.
\newblock \href {https://doi.org/10.1070/PU1997v040n04ABEH000227}
  {\path{doi:10.1070/PU1997v040n04ABEH000227}}.

\bibitem{CampbellSchonfeldWingate1983}
D.~K. Campbell, J.~F. Schonfeld, C.~A. Wingate, {Resonance structure in
  kink-antikink interactions in $\ensuremath{\varphi}^4$ theory }, Physica D 9
  (1983) 1.
\newblock \href {https://doi.org/10.1016/0167-2789(83)90289-0}
  {\path{doi:10.1016/0167-2789(83)90289-0}}.

\bibitem{GoodmanHaberman2005}
R.~H. Goodman, R.~Haberman, Kink-antikink collisions in the $\phi^{4}$
  equation: The n-bounce resonance and the separatrix map, SIAM J. Appl. Dyn.
  Syst 4~(4) (2005) 1195--1228.

\bibitem{TakyiWeigel2016}
I.~Takyi, H.~Weigel, {Collective Coordinates in One-Dimensional Soliton Models
  Revisited}, Phys. Rev. D 94~(8) (2016) 085008.
\newblock \href {http://arxiv.org/abs/1609.06833} {\path{arXiv:1609.06833}},
  \href {https://doi.org/10.1103/PhysRevD.94.085008}
  {\path{doi:10.1103/PhysRevD.94.085008}}.

\bibitem{CamposMohammadi2021}
J.~a. G.~F. Campos, A.~Mohammadi, {Wobbling double sine-Gordon kinks}, JHEP 09
  (2021) 067.
\newblock \href {http://arxiv.org/abs/2103.04908} {\path{arXiv:2103.04908}},
  \href {https://doi.org/10.1007/JHEP09(2021)067}
  {\path{doi:10.1007/JHEP09(2021)067}}.

\bibitem{AlonsoIzquierdoQueirogaNunesNieto2021}
A.~Alonso~Izquierdo, J.~Queiroga-Nunes, L.~M. Nieto, {Scattering between
  wobbling kinks}, Phys. Rev. D 103~(4) (2021) 045003.
\newblock \href {http://arxiv.org/abs/2007.15517} {\path{arXiv:2007.15517}},
  \href {https://doi.org/10.1103/PhysRevD.103.045003}
  {\path{doi:10.1103/PhysRevD.103.045003}}.

\bibitem{AdamOlesRomanczukiewiczWereszczynski2019}
C.~Adam, K.~Oles, T.~Romanczukiewicz, A.~Wereszczynski, {Spectral Walls in
  Soliton Collisions}, Phys. Rev. Lett. 122~(24) (2019) 241601.
\newblock \href {http://arxiv.org/abs/1903.12100} {\path{arXiv:1903.12100}},
  \href {https://doi.org/10.1103/PhysRevLett.122.241601}
  {\path{doi:10.1103/PhysRevLett.122.241601}}.

\bibitem{EvslinHalcrowRomanczukiewiczWereszczynski2022}
J.~Evslin, C.~Halcrow, T.~Romanczukiewicz, A.~Wereszczynski, {Spectral walls at
  one loop}, Phys. Rev. D 105~(12) (2022) 125002.
\newblock \href {http://arxiv.org/abs/2202.08249} {\path{arXiv:2202.08249}},
  \href {https://doi.org/10.1103/PhysRevD.105.125002}
  {\path{doi:10.1103/PhysRevD.105.125002}}.

\bibitem{ChristLee1975}
N.~H. Christ, T.~D. Lee, {Quantum Expansion of Soliton Solutions}, Phys. Rev. D
  12 (1975) 1606.
\newblock \href {https://doi.org/10.1103/PhysRevD.12.1606}
  {\path{doi:10.1103/PhysRevD.12.1606}}.

\bibitem{TrullingerFlesch1987}
S.~E. Trullinger, R.~J. Flesch, {Parent potentials for an infinite class of
  reflectionless kinks}, J. Math. Phys. 28 (1987) 1683--1690.
\newblock \href {https://doi.org/10.1063/1.527476}
  {\path{doi:10.1063/1.527476}}.

\bibitem{CasahorranNam1991a}
J.~Casahorran, S.~Nam, {Kinks and bounces from zero modes}, Int. J. Mod. Phys.
  A 6 (1991) 5467--5480.
\newblock \href {https://doi.org/10.1142/S0217751X91002574}
  {\path{doi:10.1142/S0217751X91002574}}.

\bibitem{FloresHidalgoSvaiter2002}
G.~Flores-Hidalgo, N.~F. Svaiter, {Constructing bidimensional scalar field
  theory models from zero mode fluctuations}, Phys. Rev. D 66 (2002) 025031.
\newblock \href {https://doi.org/10.1103/PhysRevD.66.025031}
  {\path{doi:10.1103/PhysRevD.66.025031}}.

\bibitem{BoyaCasahorran1990}
L.~J. Boya, J.~Casahorran, {Quantum masses for a general family of
  bidimensional kinks}, Phys. Rev. D 41 (1990) 1342.
\newblock \href {https://doi.org/10.1103/PhysRevD.41.1342}
  {\path{doi:10.1103/PhysRevD.41.1342}}.

\bibitem{AlonsoIzquierdoMateosGuilarte2012}
A.~Alonso-Izquierdo, J.~Mateos~Guilarte, {On a family of (1+1)-dimensional
  scalar field theory models: kinks, stability, one-loop mass shifts}, Annals
  Phys. 327 (2012) 2251--2274.
\newblock \href {http://arxiv.org/abs/1205.3069} {\path{arXiv:1205.3069}},
  \href {https://doi.org/10.1016/j.aop.2012.04.014}
  {\path{doi:10.1016/j.aop.2012.04.014}}.

\bibitem{Zhong2022a}
Y.~Zhong, {Singular P\"oschl-Teller II potentials and gravitating kinks}, JHEP
  09 (2022) 165.
\newblock \href {http://arxiv.org/abs/2207.12681} {\path{arXiv:2207.12681}},
  \href {https://doi.org/10.1007/JHEP09(2022)165}
  {\path{doi:10.1007/JHEP09(2022)165}}.

\bibitem{Stoetzel1995}
B.~Stoetzel, {Two-dimensional gravitation and Sine-Gordon solitons}, Phys. Rev.
  D 52 (1995) 2192--2201.
\newblock \href {http://arxiv.org/abs/gr-qc/9501033}
  {\path{arXiv:gr-qc/9501033}}, \href
  {https://doi.org/10.1103/PhysRevD.52.2192}
  {\path{doi:10.1103/PhysRevD.52.2192}}.

\bibitem{Zhong2021}
Y.~Zhong, {Revisit on two-dimensional self-gravitating kinks: superpotential
  formalism and linear stability}, JHEP 04 (2021) 118.
\newblock \href {http://arxiv.org/abs/2101.10928} {\path{arXiv:2101.10928}},
  \href {https://doi.org/10.1007/JHEP04(2021)118}
  {\path{doi:10.1007/JHEP04(2021)118}}.

\bibitem{ZhongLiLiu2021}
Y.~Zhong, F.-Y. Li, X.-D. Liu, {K-field kinks in two-dimensional dilaton
  gravity}, Phys. Lett. B 822 (2021) 136716.
\newblock \href {http://arxiv.org/abs/2108.10166} {\path{arXiv:2108.10166}},
  \href {https://doi.org/10.1016/j.physletb.2021.136716}
  {\path{doi:10.1016/j.physletb.2021.136716}}.

\bibitem{FengZhong2022}
J.~Feng, Y.~Zhong, {Scalar perturbation of gravitating double-kink solutions},
  EPL 137~(4) (2022) 49001.
\newblock \href {http://arxiv.org/abs/2202.02946} {\path{arXiv:2202.02946}},
  \href {https://doi.org/10.1209/0295-5075/ac56ae}
  {\path{doi:10.1209/0295-5075/ac56ae}}.

\bibitem{ZhongGuoLiu2023}
Y.~Zhong, H.~Guo, Y.-X. Liu, {Kink solutions in generalized 2D dilaton
  gravity}, Phys. Lett. B 849 (2023) 138471.
\newblock \href {http://arxiv.org/abs/2308.13786} {\path{arXiv:2308.13786}},
  \href {https://doi.org/10.1016/j.physletb.2024.138471}
  {\path{doi:10.1016/j.physletb.2024.138471}}.

\bibitem{WangZhongWang2024}
Z.~Wang, Y.~Zhong, H.~Wang, {Gravitating kinks with asymptotically flat
  metrics}, EPL 146~(5) (2024) 59001.
\newblock \href {http://arxiv.org/abs/2402.05486} {\path{arXiv:2402.05486}},
  \href {https://doi.org/10.1209/0295-5075/ad49d0}
  {\path{doi:10.1209/0295-5075/ad49d0}}.

\bibitem{RandallSundrum1999}
L.~Randall, R.~Sundrum, A large mass hierarchy from a small extra dimension,
  Phys. Rev. Lett. 83 (1999) 3370--3373.
\newblock \href {http://arxiv.org/abs/hep-ph/9905221}
  {\path{arXiv:hep-ph/9905221}}, \href
  {https://doi.org/10.1103/PhysRevLett.83.3370}
  {\path{doi:10.1103/PhysRevLett.83.3370}}.

\bibitem{RandallSundrum1999a}
L.~Randall, R.~Sundrum, An alternative to compactification, Phys. Rev. Lett. 83
  (1999) 4690--4693.
\newblock \href {http://arxiv.org/abs/hep-th/9906064}
  {\path{arXiv:hep-th/9906064}}, \href
  {https://doi.org/10.1103/PhysRevLett.83.4690}
  {\path{doi:10.1103/PhysRevLett.83.4690}}.

\bibitem{CasahorranNam1991}
J.~Casahorran, S.~Nam, {Singular superpotentials and explicit breaking of
  supersymmetry}, Int. J. Mod. Phys. A 6 (1991) 2729--2742.
\newblock \href {https://doi.org/10.1142/S0217751X91001325}
  {\path{doi:10.1142/S0217751X91001325}}.

\bibitem{GangopadhyayaMallowRasinariu2017}
A.~Gangopadhyaya, J.~V. Mallow, C.~Rasinariu, {Supersymmetric Quantum
  Mechanics}: {An Introduction}, World Scientific, 2017.
\newblock \href {https://doi.org/10.1142/10475} {\path{doi:10.1142/10475}}.

\bibitem{OgundipeEvslinZhangGuo2024}
K.~Ogundipe, J.~Evslin, B.~Zhang, H.~Guo, {A (2+1)-dimensional domain wall at
  one-loop}, JHEP 05 (2024) 098.
\newblock \href {http://arxiv.org/abs/2403.14062} {\path{arXiv:2403.14062}},
  \href {https://doi.org/10.1007/JHEP05(2024)098}
  {\path{doi:10.1007/JHEP05(2024)098}}.

\bibitem{OgundipeEvslin2024}
K.~Ogundipe, J.~Evslin, {Perturbative approach to time-dependent quantum
  solitons}, JHEP 06 (2024) 174.
\newblock \href {http://arxiv.org/abs/2403.13232} {\path{arXiv:2403.13232}},
  \href {https://doi.org/10.1007/JHEP06(2024)174}
  {\path{doi:10.1007/JHEP06(2024)174}}.

\bibitem{EvslinLiu2024}
J.~Evslin, H.~Liu, {Reflection coefficient of a reflectionless kink}, Phys.
  Rev. D 109~(8) (2024) 085019.
\newblock \href {http://arxiv.org/abs/2402.17968} {\path{arXiv:2402.17968}},
  \href {https://doi.org/10.1103/PhysRevD.109.085019}
  {\path{doi:10.1103/PhysRevD.109.085019}}.

\bibitem{EvslinLiu2024a}
J.~Evslin, H.~Liu, {Elastic Kink-Meson scattering}, JHEP 04 (2024) 072.
\newblock \href {http://arxiv.org/abs/2311.14369} {\path{arXiv:2311.14369}},
  \href {https://doi.org/10.1007/JHEP04(2024)072}
  {\path{doi:10.1007/JHEP04(2024)072}}.

\bibitem{EvslinLiu2023}
J.~Evslin, H.~Liu, {(Anti-)Stokes scattering on kinks}, JHEP 03 (2023) 095.
\newblock \href {http://arxiv.org/abs/2301.04099} {\path{arXiv:2301.04099}},
  \href {https://doi.org/10.1007/JHEP03(2023)095}
  {\path{doi:10.1007/JHEP03(2023)095}}.

\bibitem{EvslinRoystonZhang2023}
J.~Evslin, A.~B. Royston, B.~Zhang, {Cut-off kinks}, JHEP 01 (2023) 073.
\newblock \href {http://arxiv.org/abs/2210.16523} {\path{arXiv:2210.16523}},
  \href {https://doi.org/10.1007/JHEP01(2023)073}
  {\path{doi:10.1007/JHEP01(2023)073}}.

\bibitem{Evslin2022}
J.~Evslin, {Moving kinks and their wave packets}, Phys. Rev. D 105~(10) (2022)
  105001.
\newblock \href {http://arxiv.org/abs/2202.04905} {\path{arXiv:2202.04905}},
  \href {https://doi.org/10.1103/PhysRevD.105.105001}
  {\path{doi:10.1103/PhysRevD.105.105001}}.

\bibitem{Evslin2021}
J.~Evslin, {The two-loop $\phi^4$ kink mass}, Phys. Lett. B 822 (2021) 136628.
\newblock \href {http://arxiv.org/abs/2109.05852} {\path{arXiv:2109.05852}},
  \href {https://doi.org/10.1016/j.physletb.2021.136628}
  {\path{doi:10.1016/j.physletb.2021.136628}}.

\bibitem{GuoEvslin2020}
H.~Guo, J.~Evslin, {Finite derivation of the one-loop sine-Gordon soliton
  mass}, JHEP 02 (2020) 140.
\newblock \href {http://arxiv.org/abs/1912.08507} {\path{arXiv:1912.08507}},
  \href {https://doi.org/10.1007/JHEP02(2020)140}
  {\path{doi:10.1007/JHEP02(2020)140}}.

\bibitem{Evslin2019}
J.~Evslin, {Manifestly Finite Derivation of the Quantum Kink Mass}, JHEP 11
  (2019) 161.
\newblock \href {http://arxiv.org/abs/1908.06710} {\path{arXiv:1908.06710}},
  \href {https://doi.org/10.1007/JHEP11(2019)161}
  {\path{doi:10.1007/JHEP11(2019)161}}.

\end{thebibliography}





\end{document}